\documentclass[11pt,a4paper]{article}

\usepackage{fancyhdr,color, graphics, epsfig}
\usepackage[T1]{fontenc}
\usepackage[latin9]{inputenc}
\usepackage{array}
\usepackage{amsmath}
\usepackage{amssymb}
\usepackage{graphicx}
\usepackage{url}
\usepackage{pifont}
\usepackage{rotating}
\usepackage{authblk}
\usepackage{hyperref}
\usepackage{url}

\title{QoE Modelling, Measurement and Prediction:
A Review}
\author[*]{Karan Mitra}
\author[**]{Arkady Zaslavsky}
\author[*]{Christer {\AA}hlund}
\affil[*]{ Lule{\aa} University of Technology, Skellefte{\aa}, Sweden}
\affil[* ]{\{karan.mitra, christer.ahlund\}@ltu.se}
\affil[**]{CSIRO, Canberra, Australia}
\affil[** ]{arkady.zaslavsky@csiro.au}
\begin{document}

\maketitle

\begin{abstract}
In mobile computing systems, users can access network services anywhere and anytime using mobile devices such as tablets and smart phones. These devices connect to the Internet via network or telecommunications operators. Users usually have some expectations about the services provided to them by different operators. Users' expectations along with additional factors such as cognitive and behavioural states, cost, and network quality of service (QoS) may determine their quality of experience (QoE). If users are not satisfied with their QoE, they may switch to different providers or may stop using a particular application or service. Thus, QoE measurement and prediction techniques may benefit users in availing personalized services from service providers. On the other hand, it can help service providers to achieve lower user-operator switchover.
This paper presents an extensive review of the state-the-art research in the area of QoE modelling, measurement and prediction. In particular, we investigate and discuss the strengths and shortcomings of existing techniques. Finally, we present future research directions for developing novel QoE measurement and prediction techniques.
\end{abstract}

\section{Introduction}

Mobile devices connect to the Internet via one or more telecommunications
operators. Users usually have expectations about the services
they receive from these operators \cite{sung,mitraicme2011}. Based
on their experience about the services they receive, they either choose
to stay with the current operator or they switch to a new operator.
For example, in 2011, Vodafone Australia nearly lost 440,000 customers to different operators such as Telstra and Optus%
\footnote{http://www.itnews.com.au/News/290168,vodafone-australia -churn-nears-half-a-million-for-2011.aspx.
Retrieved 12/07/12.%
}. Telecommunications operators are interested in maximizing their
revenue by trying to retain their customers. On the other hand, users consider operators that meet their expectations based on cost and quality of service (QoS) offered to them. 

It is widely assumed \cite{nokia,jain,1631338,Jumisko2008} that by maximizing network QoS (e.g., increasing network bandwidth
and/or increasing wireless signal strength) or by reducing the cost of
services, users will be satisfied with the services provided to them.
On the other hand, \cite{sung,kilkki} argue that users may or may not be satisfied with QoS and the cost
of services offered to them by operators.
For example, consider a statement posted on the Apple forum:%
\footnote{https://discussions.apple.com/thread/3437795?start=180\& tstart=0. 
Retrieved 11/07/12.%
} 

\textbf{Example:} \emph{{}``I am having the same issues as everyone else...Phone shows
5 bars on \textquotedbl{}4G.\textquotedbl{} Makes calls and texts
just fine but no imessage or internet (safari as well as any other
apps that require connectivity). Right now the two things I have noticed
are that I'm more likely to have it work late at night (11pm-2am)
and more likely to have it work when I'm outdoors. Most of my problems
are experienced while at work (9am-6pm) and when I leave work it tends
to work for a while (lunch break, errands) but not always. I did try
turning off all the 'System Services' under 'Location services' and
have not noticed any difference in my phone's behavior.''} 

This example shows that positive user experience may not be guaranteed even with 4G networks.
 There is a need to understand users' perception of services or their Quality of Experience (QoE) \cite{jain,kilkki,sung}. QoE
as a term is often misunderstood and is narrowly associated with QoS
\cite{kilkki}. We argue that QoE is a multidisciplinary and a multidimensional concept. It involves concepts from several fields including human computer interaction, cognitive and behavioral science, computer networks and economics \cite{kilkki,sung,essay,marez2007}. 

ITU-T \cite{ituqoe} defines QoE as: \textit{``The overall acceptability
of an application or service, as perceived subjectively by the end-user.''} 
 It's worth noting that ITU-T also consider the following statements:
\textit{``Quality of Experience includes the complete end-to-end
system effects (client, terminal, network, services infrastructure,
etc.)''} and `\textit{`Overall acceptability may be influenced by
user expectations and context.''} The key point to note here is that ITU-T does not define
what it means by \emph{{}``context''} and how experts can measure
users' \emph{{}``expectations''}. 

\emph{Context} is any information that assists in determining a situation(s) related
to a user, network or device \cite{Dey}. For example, from GPS coordinates,
user-related situation can be inferred as: \textquotedblleft{}user is at work\textquotedblright{}.
From delay of 50 ms and packet losses of 0\%, network-related situation
can be determined as \textquotedblleft{}network is not congested\textquotedblright{}.
We consider context as any information that assists in determining users' QoE. Context can be static and dynamic.  Static context does
not change often, while dynamic context changes over a period of time
and is difficult to predict. Static context may include user's application 
preferences, their security requirements and cost.

In real-life environments, context can be
highly dynamic and stochastic i.e., it can change in a very short period
of time and is uncertain; it can be imperfect; it can exhibit a range
of temporal characteristics; it can have several alternative representations; it
can be interrelated; it can be distributed; and it may not be available at
a particular time. The timely collection and processing of context may be crucial 
as it may loose its accuracy. Dynamic context may include
user location, velocity, network load, battery power,
memory/CPU utilization, presence and SNR.   

ETSI \cite{ETSI2010} defines QoE as: \textit{``A measure of user
performance based on both objective and subjective psychological measures
of using an ICT service or product.''} It also highlights
the importance of technical parameters such as those related to QoS
and communication context.

Nokia  \cite{nokia} defines QoE as: \textit{``QoE is how a user perceives the usability
of a service when in use - how satisfied he or she is with a service. The term QoE
refers to the perception of the user about the quality of a particular service or network.''}

Mitra, Zaslavsky and {\AA}hlund \cite{mitraTMC} define QoE as: 
\emph{{}``Quality of experience (QoE) is a metric that depends on
the underlying QoS along with a person's preferences towards a particular
object or service where his/her preferences are defined by his/her
personal attributes related to expectations, experiences, behaviour,
cognitive abilities, object's attributes and the environment surrounding
that person''}.


\begin{table}
\begin{centering}
\caption{Context parameters related to user, application, device and network
environment that need to be considered when modelling, measuring and
predicting QoE.}

\par\end{centering}

\centering{}%
\begin{tabular}{|p{3cm}|p{4.6cm}|}
\hline 
Context classes & Context parameters\tabularnewline
\hline 
\hline 
User and user environment & location, temperature, heart rate, eye movement, amount of sweat,
social context, people nearby, light, background noise, age, gender\tabularnewline
\hline 
Tool/device/object & screen size, design layout, resolution, general intuitiveness, buttons
placement, input/output methods, appeal, usability\tabularnewline
\hline 
Application & type, requirements\tabularnewline
\hline 
Network & type, bandwidth,delay, jitter, packet loss, RTT, loss burst size,
protocols used, received signal strength, congestion levels\tabularnewline
\hline 
\end{tabular}
\end{table}

QoE can be computed using subjective and objective test. Subjective tests require direct data collection from end users. These tests lead to higher costs in terms of time and money.  On the other hand, objective test directly predict users' QoE without requiring subjective tests. For example, the ITU-T E-Model \cite{G.107}
considers QoS parameters (e.g., network delay and packet losses) to
compute QoE in terms of the mean opinion score (MOS).

In table 1 we enlist several context parameters related to application, device, network and the user environment that may assist in computing users' QoE. Along with context, there can be a plethora of QoE parameters such as  enjoyment, user satisfaction, technology acceptance, efficiency, accuracy and perceived ease-of-use \cite{brooks2010,1631338,5246986}.  Studying and modelling these parameters to determine QoE is a challenging task \cite{brooks2010,1631338,sung,mitrasac2011}. There can be inter-dependencies and non-linear relationships 
between context and QoE parameters \cite{correlation,mitraTMC}. Further, some parameters may be hidden. By the term \textit{"hidden"} we mean that some parameters may not be observed directly. Thus, these parameters may be  hard to measure and quantify.    QoE modelling and measurement may require the combination of several QoE parameters to determine  overall QoE. For example, combining QoE parameters such as ``user satisfaction'' and ``technology acceptance'' to compute users' overall QoE. This problem can be aggravated by the fact that each QoE parameter can be measured on a different scale or by considering different units of measurement \cite{brooks2010,mitraicme2011}. For example, ``user satisfaction" can be measured on the scale of 1 to 5. On the other hand, ``technology acceptance" can be measured using simple ``yes" or ``no". 

The contribution of this  paper is to review research pertaining to QoE modelling, measurement and prediction.  In doing so, we identify and highlight several important challenges that should be addressed to realize efficient techniques for QoE modelling, measurement and prediction. This paper is organized
as follows: Section II presents an in-depth discussion on subjective
and objective methods and introduces the research problems pertaining to QoE modelling, measurement and prediction. Sections III and IV presents the state-of-the-art methods and discusses their advantages and shortcomings. 
Section V presents the future research directions that should be pursued to
realize efficient methods for QoE modelling, measurement and prediction.
Finally, section VI presents the conclusion. 

%

\section{Methods for QoE Measurement and Prediction}

\subsection{Subjective Methods}

QoE measurement can be performed using subjective and objective tests \cite{brooks2010,moormonet,Li-yuan2006,1631338,takahashi}. Subjective
tests involve direct data collection from users. For example, in
the form of user ratings. Standardization bodies such as the ITU-T in its ITU-T P.800 recommendation \cite{P.800} presents a methodology for conducting subjective tests. This recommendation also defines a method to measure users' QoE based on a score called the mean opinion score (MOS) \cite{P.800}. MOS is used widely for subjective voice/video quality assessment where human test subjects grade their overall experience on the Absolute Category Rating Scale (ACR). This scale typically comprises of five alternatives, for example,  '5' means {}``excellent'',
'4' means {}``very good'', '3', '2' and '1' means {}``good'', ``fair'' and ``poor'' respectively.

There are several problems that arise while conducting subjective tests. For example, a large sample space is required to get credible results \cite{rix}. These tests can be expensive and time consuming  \cite{takahashi,brooks2010,mollericc2009}. Hence,
subjective tests are mainly limited to major telecommunication providers.
Further, the native language of human subjects might not be same across test subjects and the results obtained via subjective tests can be biased or even incomplete \cite{Knoche99}. 

Several researchers \cite{madm,Janowski2009,brooks2010,mitraicme2011} also noted problems while adhering to the ITU-T P.800 recommendation for conducting subjective tests. The biggest problem with MOS is that an average of user ratings is computed.  Mathematical operations such as computing mean and standard deviation cannot be applied on subjective ratings as these ratings are categorical in nature (e.g., "excellent" and "fair"). The human test subjects ranks the alternatives  on the categorical scale where the distance between these alternatives  cannot be known \cite{madm,Janowski2009,brooks2010,mitraicme2011}. Hence, mathematical operations cannot be applied.  Nonetheless, MOS is the most widely used method to assess subjective ratings in both industry and academia. 

\subsection{Objective Methods}

Takahashi \emph{et al. }\cite{takahashi}\emph{ }argued for developing
objective methods to estimate QoE for multimedia applications.
Objective methods such as the ITU-T E-Model \cite{G.107}, PESQ \cite{pesq},
PSQA \cite{psqa}, USI \cite{usi}, once
developed, can be used for \textit{QoE prediction} without requiring subjective
tests. Most of objective methods map their scores onto the MOS to determine QoE. For example, the ITU-T E-Model \cite{G.107} computes
the R-factor in range of {[}0:100{]} for narrowband codecs and {[}0:129]
for wideband codecs. The R-factor is then mapped to the MOS using a non-linear equation to determine users' QoE between the range of 1 and 5.

Objective methods are hard to develop, model and deploy due to large parameter space.
Further, any modification made to current objective methods by addition or deletion of parameters may require new tests to fine-tune the current model or to derive new statistical models for QoE prediction \cite{brooks2010}. We assert that the current objective models
such as \cite{G.107}, \cite{lingfen2006} and \cite{fiedlernetwork}
are based on simplistic assumptions regarding QoE prediction. For example,
Fiedler and Hossfeld \cite{fiedlernetwork} considers only one to two QoS
parameters to predict QoE.

\subsection*{The role of context}

Most of the QoE measurement and prediction methods were developed under controlled laboratory environments with a limited number of objective and subjective parameters such as delay, jitter, packet loss, and bandwidth. Moor \emph{et al.} \cite{moormonet},
Jumisko-Pyykk\"{o} and Hannuksela \cite{Jumisko2008}, Ickin \emph{et al.} \cite{ickin12} and Mitra \emph{et al. }\cite{mitraicme2011,mitradbn}\emph{
}argued for QoE measurement and prediction in real-life user environments.
Jumisko-Pyykk\"{o} and Hannuksela \cite{Jumisko2008} shows that users'
QoE differs in laboratory and real-life user environments. In real-life environments, context can change dynamically while users
are on-the-move. For example, at different user locations, QoS can
vary. Factors such as time-of-the-day, can help explain rise in network congestion (see example in section I) leading to decrease in users' QoE. Further, users' social context changes
throughout the day leading to variation in QoE. For example, users'
QoE may be affected if there are people nearby. Nearly all objective
methods  (e.g., \cite{usi,fiedlernetwork,oneclick,psqa}) developed till date do not consider grouping of several context parameters
such as user location, time of the day and screen resolution for QoE prediction. 
Thus, objective methods not considering context may not provide accurate
QoE predictions in mobile computing systems. \textbf{}

Brooks and Hestnes \cite{brooks2010} discussed the importance of
considering both subjective and objective methods for QoE measurement
and prediction. They also discussed the importance of measuring QoE
on a single scale by combining several QoE parameters. Moor \emph{et al.} \cite{moormonet} and Mitra \emph{et
al.} \cite{mitraicme2011,mitrasac2011} also suggested the combination
of both subjective and objective methods for QoE measurement. We conclude
that there can be a plethora of context and QoS parameters affecting
QoE. In fact, different QoE parameters can also affect each other as shown
in Fig 1. Further, QoE parameters can be measured on a different scale
\cite{brooks2010} as shown in Fig 2.

 We assert that there is a need to develop methods that correctly identify and model these parameters in order to measure and predict QoE on a single scale. Once developed, these methods may benefit both telecommunication operators and end users. For example, objective  methods may be used to provide users with personalized services on their mobile devices. On the other hand, operators may be able to minimize network churn. 
 \begin{figure}
 
 \begin{centering}
 \includegraphics[scale=0.3008]{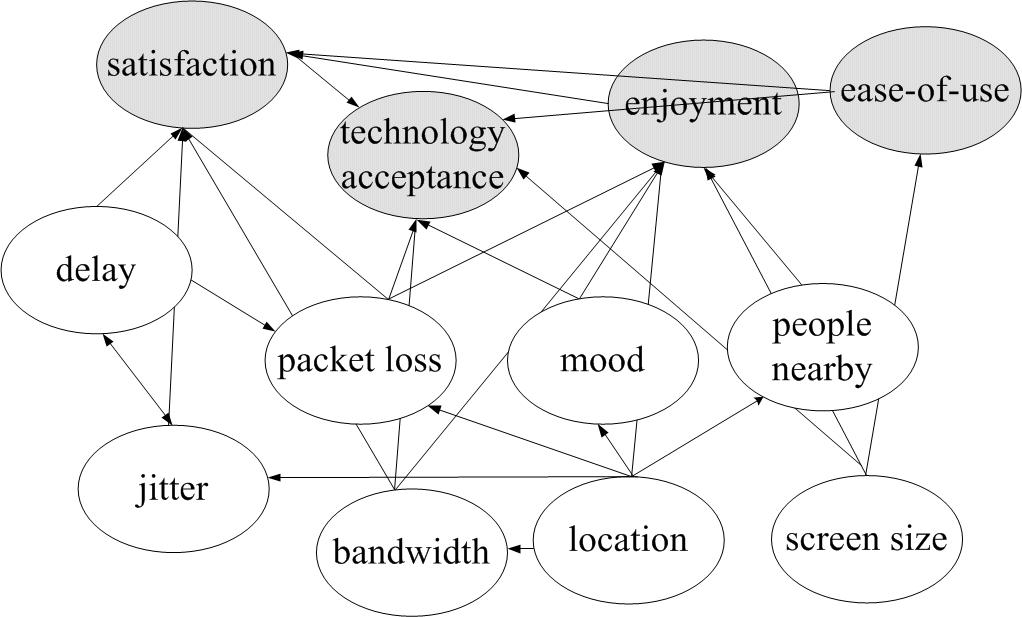}\caption{Parameter relationships between context and QoE parameters can
 be complex. Grey ovals depict QoE parameters. White ovals depicts context parameters.}
 \par\end{centering}
 \end{figure}

\subsection{Research Challenges}
Researchers
considering the problem of QoE modelling, measurement and prediction face a number of researcher challenges. These include:
\begin{enumerate}
\item \textbf{QoE modelling:} QoE measurement and prediction may involve
a large parameter space comprising of several QoE and context parameters
\cite{5246986} as shown in Fig. 1. There can be \emph{N} context parameters affecting \emph{M} QoE parameters. Further \emph{M}
QoE parameters can affect each other. Thus, selecting relevant parameters and defining and
finding relationships between these parameters can be challenging.
The relationships  between these parameters are usually non-linear and hard to quantify. This necessitates the development of novel QoE modelling techniques 
to model all these parameters efficiently.  The QoE models should not
only be conceptual, but should also transcend to solving the challenges associated with QoE measurement and prediction. For example, rather than simply classifying and representing
the parameters, QoE models should directly be used for QoE measurement 
and prediction.

\item \textbf{QoE measurement and prediction:}
The challenge of QoE measurement and prediction involving multiple QoE and context parameters is not well addressed.  Consider Fig. 2, each QoE parameter can be measured on a different scale and may involve different units of measurement \cite{brooks2010,mitraicme2011}. These scales can be qualitative or quantitative. For example, QoE parameter ``user satisfaction'' is measured using an ordinal (qualitative) scale  involving ratings 1 to 5 (see Fig 2b). On the other hand, QoE parameter  ``technology acceptance'' may  not require a scale as it can be measured using a simple {}``yes'' or {}``no''   type questions \cite{1631338}. Thus, determining QoE based on different types of scale and/or different units of measurement remains a challenging task. 
Further, current methods do not explicitly deal with the problem of imprecision in QoE measurement and prediction due to uncertainty caused by scarce and sparse data or other uncontrollable factors prevalent in both laboratory and real-life environments \cite{mitrasac2011}. 

\item \textbf{QoE measurement and prediction over time:} \cite{mitradbn,karapanos2010,perkis} argue that QoE evolves over time. By repeated use of a service, a user may or may not be satisfied by their QoE. Thus, QoE measurement and prediction at a single point in time may not yield correct results and may have to be done over a longer period time. This necessitates the development of novel techniques for QoE modelling, measurement and prediction over time \cite{mitradbn}.

\end{enumerate}
In the following sections, we discuss these challenges in detail and discuss the state-of-the-art research and present future research directions that may lead to efficient methods for  QoE modelling, measurement and prediction.

\begin{figure}

\begin{centering}
\includegraphics[scale=0.50]{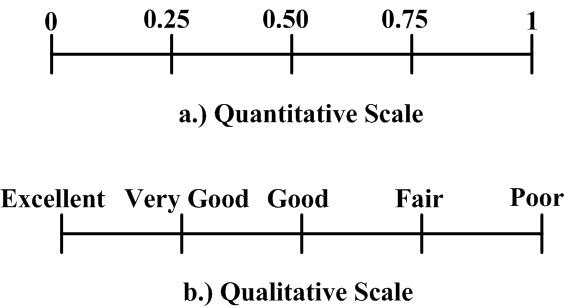}\caption{Typical scales for QoE measurement.}
\par\end{centering}
\end{figure}

\section{QoE Modelling }

Perkis \emph{et al.} \cite{perkis} presented a conceptual model for QoE measurement. Their model includes technology and user related parameters to measure QoE. The authors classified these parameters as either quantifiable or unquantifiable.  Quantifiable parameters  include bandwidth, delay and jitter. Parameters such as expectation, attitude, ease-of-use are related to user and are deemed to be unquantifiable. We differ with authors here that some user related parameters such as ease-of-use can be quantified. 

The biggest problem with their model is that parameters can only be classified and represented as a tree structure and it cannot be used to measure and predict QoE. Further, their model assumes that all parameters are independent of each other. In reality, this might not be the case. For example, QoE parameter, ``user satisfaction'' may affect another QoE parameter such as ``technology acceptance'' which may determine whether a user accepts a particular technology or not. 

Sun \cite{sung} developed a conceptual QoE model for multimedia applications such as video-on-demand. Their model was inspired by the customer satisfaction model, the disconfirmation of expectations model (DEM) and the sport spectator satisfaction index. Their model was based on the premise that users have expectations (both positive and negative) and their expectations are based on their needs. If their perceptions about a service are met, the users might have positive disconfirmation, leading them being satisfied. On the other hand, if the users have a negative disconfirmation i.e., if their expectations are not met, they might not be satisfied. The author modelled users' perception as a function of affective and cognitive responses. The author then used these responses to predict the overall QoE by deriving utility functions in the context of mobile video steaming applications. The author, however, did not present a method to integrate new context and QoE parameters and assumes independence between parameters.

Gong \emph{et al.} \cite{gongqoemodel} proposed a pentagram model for QoE measurement. The main highlight of their model  is that it combines several QoE parameters i.e., service availability, usability, retainability, integrity and instantaneousness to determine a single QoE value. Each QoE parameter is a function (linear or ratio) of one or more QoS parameters. For example, service integrity is the function of delay, jitter and packet loss ratio. Each QoE parameter is then represented in a pentagram.  The computed QoE value is mapped to the normalized MOS scale to determine a single QoE value. However, the author do not discuss how their model can be extended with new context and QoE parameters. Further, their model does not define dependencies between context parameters i.e, all the parameters are considered independent. The main highlight of their model is that it is practical and can be used in real applications.

A comprehensive treatment of QoE modelling problem was considered by Wu \emph{et al.} \cite{1631338}. The authors presented a conceptual QoE model comprising of QoE and QoS constructs. The QoS constructs include factors such as interactivity, vividness and consistency. These factors in turn describe network QoS parameters. For instance, interactivity depends on delay and vividness relates to metrics such peak-signal-to-noise-ratio (PSNR). The QoE construct consists of parameters such as concentration, attention and technology acceptance. The authors also presented a theoretical framework for QoE measurement. The quantitative  mappings between QoE and QoS were established via correlation analysis. However, all parameters affecting QoE were considered independent.

QoE models \cite{sung,gongqoemodel,1631338} are  mainly limited to QoS parameters. These models do not consider other context parameters such as  location, type of mobile device, as time-of-the day, etc. (as mentioned in table 1). Mitra \textit{et al}. \cite{mitrasac2011,mitraicme2011} argue that inclusion of several context parameters in a QoE model, may lead to increase in QoE measurement and prediction accuracy especially in users' real-life  environments.

Korhonen \textit{et al.} \cite{Korhonen2010}
discussed the need for context-awareness for QoE measurement. The main contribution
of the paper is to categorize context into eight different categories
and to find \emph{triggering context } i.e.,  to find the most important
context category. Context categories identified by the authors were: environment context, person
context, device context, task context, social context, spatio-temporal
context, service and access network context. For results analysis,
a questionnaire was prepared based on the aforementioned context categories. The authors used these categorizes to analyze 
users' phone usage experience. However, the authors did not develop a
context model to reason about context to measure and predict QoE.
Further, they did not describe how their approach can be incorporated
in applications.

Laghari and Connelly \cite{laghari12} proposed
a conceptual QoE model. Their model consists of human, context, business and technology domains. However, authors simply classify parameters related to each domain.
For example, GPS related to context domain and age and gender related to human domain. The authors do not present methods whereby their conceptual model can be used to predict users' QoE. Thus, their model is limited to parameter classification.

Marez and Moor \cite{marez2007}, developed a conceptual
framework for QoE measurement. Their model consists of five components.
These include: 1) quality of effectiveness: it includes QoS parameters such as jitter, packet loss, reliability, CPU
usage, etc; 2) usability: it includes parameters such as ease-of-use.
3) quality of efficiency: it includes subjective parameters related
to device, application and network. These parameters include, user
satisfaction, speed, interface, etc; 4) expectations: as the name
suggests, this component quantify degree of users' exceptions
that are met; and 5) context: including information such as, environmental,
personal, social, cultural, technological and organizational. As with
models presented above, their model simply classify several
parameters affecting QoE. Their model does not include methods to measure and predict QoE. 

Mitra \textit{et. al}  \cite{mitraicme2011, mitraTMC} presented a context-aware approach called CaQoEM---Context-aware QoE Modelling and Measurement to model, measure and predict users' QoE. Their model is based on Bayesian networks (BNs) \cite{russelandnorvig} and the context spaces model \cite{Padovitz2004a}. By using BNs, the relationships between context and QoE parameters and the relationships
among QoE parameters can be determined in a simplified and in an efficient manner. The experts simply need to define mappings \emph{casually} by linking \emph{causes} (e.g., context parameters) to \emph{effects} (e.g., QoE parameters). They do not  need to develop a precise mathematical or statistical model to determine the mappings between context and QoE parameters. The BNs can automatically handle linear and non linear relationships and can handle both discrete and continuous variables. Compared to \cite{sung,gongqoemodel,1631338} their model is capable of graceful addition and removal context and QoE parameters. It incorporates domain/experts knowledge for QoE measurement and prediction. Finally, it maps several QoE and context parameters to measure and predict users' QoE on a single scale. 

QoE models presented in \cite{perkis,sung,marez2007,laghari12} are conceptual in nature. These models do not transcend to solving the challenges associated with QoE measurement and prediction i.e., to derive mapping(s) between multiple QoE and context parameters to measure and predict users' QoE on a single scale. These models simply enable experts to classify parameters but do not propose methods to realistically measure and predict users' QoE in both laboratory and real-life settings. For example, Laghari and Connelly' \cite{laghari12} model link multiple domains together but do not provide concrete methods to determine relationships between each domain and their related parameters. 

If experts were to use conceptual models such as \cite{perkis,sung,marez2007,laghari12}, they will have to firstly determine the parameters they require. They will then have to derive statistical or mathematical models. These models may not correspond to the original conceptual model, tremendously reducing the benefits of QoE modelling. Except for \cite{gongqoemodel,mitrasac2011, mitraicme2011, 1631338, mitraTMC}, the other QoE models cannot be used directly to measure and predict QoE. Further, except for \cite{mitrasac2011,mitraicme2011, mitraTMC}, these models may not be used directly in applications and cannot be shared among experimenters/operators.

\section{QoE Measurement and Prediction }

QoE prediction methods can mainly be classified into regression-based (linear and non-linear) methods (e.g., \cite{lingfen2006,usi,oneclick,G.107,Janowski2009,mollericc2009}),
correlation analysis-based methods \cite{1631338,correlation} and artificial intelligence (AI)- and machine learning (ML)-based methods \cite{psqa,menkovski2009,mitrasac2011,mitraicme2011,mitradbn}.
In recent years, several objective QoE prediction methods were developed
for Voice Over Internet Protocol (VoIP), Internet Protocol Television (IPTV) and several other applications including web browsing
and file transfer protocol (FTP). These include, ITU-T E-Model \cite{G.107},
 Perceptual Evaluation of Speech Quality (PESQ) \cite{pesq}, Perceptual Assessment of Speech Quality Assessment
\cite{psqa}, User Satisfaction Index \cite{usi}, OneClick \cite{oneclick},
generalized linear models (GLZ) \cite{Janowski2009}, Decision Trees-based
models \cite{menkovski2009} and CaQoEM \cite{mitraicme2011,mitrasac2011,mitradbn}. 

Chen \emph{et al.} \cite{usi} proposed the  User Satisfaction Index (USI) to predict users' QoE based on VoIP call session lengths. We believe that USI's dependence on the call session lengths to predict QoE will not hold in case of mobile computing systems. In these systems, users may be on-the-move and their devices may be prone to network related impairments such as congestion and handoffs which severely hamper QoE \cite{marshgronvall,mitrawcnc,mollericc2009}. USI is not flexible and do not consider other context parameters such as user's location, pricing and time-of-the-day. Further, it only considers one QoE parameter, ``user satisfaction'' and do not provide mechanisms to include other QoE parameters, if necessary.

Chen \emph{et al.} \cite{oneclick} presented OneClick to measure and predict QoE regarding multimedia applications such as VoIP, video  streaming and gaming. The authors developed a Possion regression equation to predict users' QoE based on user click rates. The user click rate is computed when the users click the keys on their keyboard corresponding to network QoS conditions. Authors validated OneClick using experimentation and by performing two case studies. The experimental analysis comprised of VoIP and video streaming applications but considered only three human subjects.  The case studies comprised of VoIP applications such as AIM, MSN Messenger, Skype and first-person shooter games such as Halo and Unreal Tournament. 

We assert  that OneClick should be validated with a large number of user studies and with different applications. The authors argued that OneClick can be used to predict QoE in case of unmeasurable parameters (similar to hidden parameters we discussed in section I.) such as  background noise. However, the authors do not discuss how their method can model and determin relationships between unmeasurable (hidden) and measurable  parameters. Further, the authors do not provide a method to handle inter-parameter dependencies between measurable parameters. Lastly, we conclude that OneClick cannot measure users' QoE  by considering multiple QoE parameters together to predict overall QoE.


As mentioned in section III, Wu \emph{et al.} \cite{1631338} presented a comprehensive QoE framework comprising of several QoS and QoE parameters. For results validation, the authors performed three experiments to study the effects of QoS, number of people in user's environment and the type of communication medium on QoE. For the first experiment, authors concluded that an increase in one-way delay causes distraction to users. In the second experiment, authors concluded that users' performance was slightly affected with the presence of few people near to them. However, users could still perform their tasks efficiently and with little distraction. Finally, in their last experiment, the authors concluded that the choice of audio-visual communication medium affects users' QoE. For example, visual medium performs better than audio medium. However, the mixed medium i.e., audio-visual medium performs best in terms of task completion time. 

For results analysis, the relationships between QoS and QoE parameters were found using correlation analysis. However, this approach can be impractical when there are several QoE and QoS parameters as finding correlation between each parameter is a complex task. The authors do not present methods to determine inter-parameter relationship between context to predict the overall QoE. Lastly, their model incorporates multiple QoE parameters but the authors do not present a method to determine the overall QoE by considering these parameters together.

Fiedler, Hossfeld and Tran-Gia \cite{fiedlernetwork} presented a
quantitative mapping between QoS and QoE using their IQX hypothesis.
It is based on exponential relationship between QoS
and QoE parameters. The IQX hypothesis takes as an input QoS parameters
such as packet loss and jitter (in the form of p-ordered ratio) to
determine QoE in the form of PESQ MOS for VoIP applications. The authors
show that the derived non-linear regression equation can provide an
excellent mapping between QoS parameters and MOS for VoIP application.
The authors also tested their hypothesis for QoE related to web browsing
by considering weighted session time and delivered bandwidth. The main drawback of IQX hypothesis is that it only considers
one QoS parameter to predict the corresponding QoE value. The authors
did not consider the problem of integrating additional context and QoE parameters to predict the overall QoE.

Kim \emph{et al.} \cite{correlation} proposed a method for QoE prediction based on a function of QoS parameters such as delay, jitter, packet loss and bandwidth.  Firstly, a normalized QoS value is computed based on the linear weighted sum of QoS parameters. Once the QoS value is computed, it is then used to determine QoE on the scale of one to five based on another QoE function.  However, the authors do not discuss in detail how the weights of each QoS parameters can be computed. The authors validated their method based on a simple case study (related to IPTV) and did not consider experimental analysis and/or subjective tests. This raises doubts concerning their methods applicability in real systems. Further, their method is limited to QoS parameters and treats each parameter independently. In reality, this might not be the case. Lastly, the authors did not discuss how new QoE parameters can be included in their model.

Janowski and Papir \cite{Janowski2009} considered generalized linear models (GLZ) to predict QoE. GLZ is a general form of linear regression and can deal with non-normality of data. It outputs the probability distribution of users' ratings instead of simply computing mean and standard deviation. We believe that the probability distribution can be valuable for the experts to understand the diversity of user ratings. Based on the subjective tests involving 60 users, the authors validated that GLZ can capture users' ratings and provides better understanding users' ratings based on probability distribution. 

The GLZ model, models a single QoE parameter independently. It is therefore difficult to predict the overall QoE based on multiple QoE parameters. In such a case, multiple QoE models, for each QoE parameter will need to be developed. Then a new model for overall QoE prediction can be developed based on the outputs of each QoE model.

\subsection*{Discussion}
The methods \cite{usi,oneclick,fiedlernetwork,1631338,correlation} consider statistical approaches for QoE prediction. For example, linear/non-linear regression and correlation analysis. These methods involve mathematical operations such as computing average, variance and standard deviation of users' ratings. It is worth noting that the users' subjective ratings are mainly based on the ordinal scale. The ordinal scale is a rank ordered scale with a finite set of alternatives. For example, ``excellent'', ``very good'', ``good'', ``fair'' and ``poor'' \cite{madm}.  These alternatives do not express precise numerical values. Further, the distance between alternatives can not be established \cite{madm,mitraicme2011,muordinal}. Thus, we assert that mathematical operations such as mean and standard deviation cannot be applied \cite{madm,muordinal,Janowski2009,mitraicme2011}. Consequently, the methods involving MOS as a metric will also be incorrect \cite{madm,muordinal,mitraicme2011}.  

Mu \emph{et al.} \cite{muordinal} show that in case of subjective tests, normality of collected data (user ratings) cannot be verified. Further, due to the subjective nature of users' ratings, parametric statistical models cannot be applied for QoE measurement and prediction. Hence, techniques involving ordinary regression analysis will also be invalid. The authors point out that the conditions for valid statistical tests are rarely verified and documented by computer science researchers. 

Mitra \textit{et al.}\cite{mitraicme2011,mitraTMC} proposed to use a biploar interval scale \cite{madm} to map users' ratings into an interval scale (see Fig 2(a)). For example, a 5-point ordinal scale (see Fig 2(b)) is calibrated in such a manner that the best alternative, for example, ``excellent'' is assigned a maximum value, '1'; the worst alternative on the other hand is assigned the lowest value, '0'. The mid-point is also used for calibration. For example, ``good'' is assigned a value of 0.50. This means that values lower than 0.50 are less favourable compared to values higher than 0.50.  For example, a value between 0.8750 and 1 is considered to be ``excellent'' while a value in the range of 0 and 0.1250 is considered as ``poor''. This way normalized values can be used to determine a QoE rating. Thus, a bipolar scale enables an expert to perform mathematical operations such as computing mean and standard deviation and the application of parametric statistical models.

\subsection*{Artificial Intelligence and Machine Learning Based Methods}

Recently, researchers considered AI- and ML-based methods to predict users' QoE. The techniques such as decision trees (DTs), random neural networks (RNNs), hidden Markov models (HMMs), Bayesian networks (BNs) and dynamic Bayesian networks (DBNs) were applied successfully to predict users' QoE in both laboratory and real-life environments. The main reason for the success of AI- and ML-based methods may be attributed to providing solid mathematical models for QoE modelling and prediction. These models are flexible and are less prone to uncertainty regarding user ratings. Further, these methods were validated using sound techniques such as cross validation \cite{russelandnorvig}. These methods are more flexible than parametric statistical models. For example, these methods may not have to adhere to strict assumptions regrading independence and normality or residuals checks. Most importantly, these methods can efficiently deal with non-linear relationships between several parameters.

Rubino, Tirilly and Varela \cite{psqa} proposed and developed a PSQA metric for QoE prediction. The PSQA metric considers numerical values such as packet loss percentage and mean loss burst size to output MOS based on RNNs. The problem with RNNs is that it requires a large number of training samples for accurate QoE prediction. This limits their ability to learn in an online manner. In contrast to RNNs, methods based on DTs \cite{menkovski2009} and BNs \cite{mitrasac2011},\cite{mitraicme2011} learn efficiently with smaller data sets.  Another problem with PSQA is its dependence on the MOS and it cannot directly predict users' subjective ratings. On the other hand, \cite{menkovski2009},\cite{mitrasac2011},\cite{mitraicme2011} can classify and predict ordinal ratings. Further RNNs cannot incorporate non-numerical context parameters  such as user's mood or location. 

Menkovski \emph{et al.} \cite{Menkovskimomm} considered several ML classifiers such as BNs, DTs and support vector machines (SVMs) to predict users' QoE. Based on experimental analysis, the authors show that DTs and rule-based systems are suitable for predicting QoE and can outperform other AI techniques such as BNs and SVMs. The authors, however, tried to classify single QoE parameter, ``QoE acceptance'' in the form of ``yes'' or ``no''. They did not consider other QoE parameters to predict users' QoE. Further, they did not discuss how new context can be included in their method. 

A recent study conducted by Mitra, \AA{}hlund and Zaslavsky \cite{mitrawcnc} show that different form of BNs such as generative and discriminative BNs can learn with scare data with high prediction accuracy of approx. 98\%. They validated their model using simulations and considered typical network impairments prevalent in mobile computing systems such as handoffs, wireless signal fading and network congestion. 

Liu, Zhou and Song \cite{Li-yuan2006} presented an approach for QoE prediction from pervasive computing point-of-view. Their approach included a hierarchical model to represent QoS parameters and their effects on QoE. They proposed a rough-set theory based approach to predict users' QoE. However, their method has several limitations. Firstly, their method can not deal with uncertainty regarding QoE measurement arising from missing user data and if there is significant variations in user ratings. Secondly, their model employs a rule-based approach where several rules need to be manually created to predict users' QoE. Thirdly, there model do not consider several QoE parameters together to predict overall QoE. Finally, the authors validated their model using a simple case study example and did not consider subjective or experimental tests

Mitra\emph{,} \AA{}hlund and Zaslavsky \cite{mitrasac2011,mitraicme2011} developed CaQoEM, a context-aware, decision-theoretic approach for QoE measurement and prediction. Their approach incorporates BNs and utility theory to measure and predict  users' QoE. CaQoEM captures uncertainty and deal with missing user ratings in an efficient and unbiased manner. It provides simplified and efficient ways to define relationships between context, QoS and QoE parameters. Further, CaQoEM incorporates graceful addition
and removal of context parameters and can incorporate domain/experts knowledge for QoE measurement and prediction. CaQoEM can map several QoE parameters and context attributes to measure users' QoE on a single scale.

The authors validated their approach using a case study, subjective and experimental tests. Compared to \cite{psqa}, \cite{Menkovskimomm} and \cite{Janowski2009}, their results show that CaQoEM is resilient to scarce and sparse data i.e.,  widely distributed user ratings. This was achieved by considering all the QoE ratings together to determine a single scalar value that ``best describes users' QoE''  rather than just selecting the most likely outcome based on highest probability. 

In case of scarce and sparce data, methods such as \cite{Menkovskimomm},\cite{psqa}  may fail as they simply consider classification accuracy as a metric for evaluation. For example, consider a case where  six users out of ten gave \textquoteright{}5\textquoteright{} and the remaining four users gave \textquoteright{}4\textquoteright{} to VoIP call quality. In this case, \cite{Menkovskimomm} will select QoE as \textquoteright{}5\textquoteright{} with probability of 0.60. This is incorrect as it ignore ratings of other four users. CaQoEM will however, find the best alternative using BNs and utility theory. Further, CaQoEM enables experts to add their expertise into the BN model to reach a single (or multiple) conclusion(s) regarding QoE which is not possible in other QoE measurement techniques such as \cite{menkovski2009,Janowski2009,psqa,usi,oneclick}.

\subsection*{Methods for QoE Measurement and Prediction Overtime}

Karapanos \emph{et al.} \cite{karapanos2010} conducted a study concerning mobile phone usage. Their results show that users perception of innovativeness increased during the first month and then remained stable. Also, users' learnability was low for the first week and then increased sharply when  they got accustomed to their mobile devices. Perkis, Munkeby and Hillestad \cite{perkis} conducted a 4 week study regarding QoE
in 3G networks. Their results show that user expectations decreased after two weeks. They also show that MOS regarding video application decreased in the last two weeks from 3.4 to 3.1 (out of 5). These results strongly suggests that users' QoE varies with time.

Hossfeld \emph{et at. }\cite{hossfeldhmm} considered the problem
of QoE prediction over time by considering Web QoE model. The authors considered page load time as the QoS parameter and considered "user satisfaction" as the QoE parameter. Their experimental results with a number of users how that hidden Markov models \cite{Rabiner1989} can model and predict users' QoE over time. However, the problem with their approach is that their HMMs considers only one QoS and QoE parameter. If experts were to incorporate more context and QoE parameters, the HMM will be harder to train and its prediction accuracy may significantly drop \cite{murphythesis}. 

Mitra, Zaslavsky and \AA{}hlund \cite{mitradbn} presented a more generic approach for QoE modelling, measurement and prediction over time by considering DBNs \cite{russelandnorvig}. Their model considers
several QoE and context parameters to measure and predict
QoE over time. The authors developed a DBN that can handle spatio-temporal context to track and predict users' QoE over time.  The authors using simulations and case studies show that DBNs can be used
efficiently to track and predict users QoE over time. However, their
method needs further subjective tests under real-life test conditions. 

We assert that the current state-of-the-art research including standards and recommendations developed by ITU-T and ETSI do not sufficiently address the problem of QoE modelling, measurement and prediction over time. QoE measurement and prediction over time  can be beneficial to experts who are interested in understanding how users interact with their services in a long run and to establish factors that may lead to network churn. Further, mobile devices may learn users' QoE over time to perform user-centric handoffs \cite{mitraatnac}. 

Table 2 presents 
the comparison of the methods for QoE modelling, measurement and prediction.

\begin{sidewaystable}
\centering
\caption{State-of-the-art in QoE modelling, measurement and prediction. }
\begin{tabular}{|p{3cm}|p{3cm}|p{3cm}|p{1.5cm}|p{2cm}|p{1.4cm}|p{0.8cm}|p{3.7cm}|}
\hline 
Paper & Domain & Technique(s) & Context-Aware & Unified QoE Model & Multiple QoE parameters & QoE over time\tabularnewline
\hline 
\hline 
Chen \emph{et al.} \cite{usi}  & VoIP & Cox regression & No & No & No & No\tabularnewline
\hline 
Chen \emph{et al}. \cite{oneclick} & Any multimedia/gaming application & Poisson regression & No & No & No & No\tabularnewline
\hline 
Wu \emph{et al.} \cite{1631338} & Any application & Correlation analysis & No & Yes & Yes & No\tabularnewline
\hline 
Fiedler Hossfeld and Tran-Gia \cite{fiedlernetwork} & VoIP and Web browsing & Exponential function & No & No & No & No\tabularnewline
\hline 
Gong \emph{et al.} \cite{gongqoemodel} & Any application & Pentagram model & No & Yes & Yes & No\tabularnewline
\hline 
Kim, Hyun and Choi \cite{correlation} & Any application & Correlation model & No & No & No & No\tabularnewline
\hline 
Liu, Zhou and Song \cite{Li-yuan2006} & Any application & Rough set theory & No & No & No & No\tabularnewline
\hline 
Janowski and Papir \cite{Janowski2009} & FTP & Generalized linear model & No & No & No & No\tabularnewline
\hline 
Rubino Tirilly and Varela \cite{psqa} & VoIP and video & Random neural networks & No & No & No & No\tabularnewline
\hline 
Menkowski \emph{et al.} \cite{Menkovskimomm} & IPTV & Decision trees & No & No & Yes & No\tabularnewline
\hline 
Mitra, \AA{}hlund and Zaslavsky \cite{mitraicme2011,mitraTMC} & Any application & Bayesian Networks and utility theory & Yes & Yes & Yes & No\tabularnewline
\hline 
Mitra, Zaslavsky and \AA{}hlund \cite{mitradbn} & Any application & Dynamic Bayesian networks and utility theory & Yes & Yes & Yes & Yes\tabularnewline
\hline 
Hossfeld \emph{et al.}\cite{hossfeldhmm} & Web application & Support vector machines, iterative exponential regressions and two-dimensional
hidden Markov models. & No & No & No & Yes\tabularnewline
\hline 
\end{tabular}

\end{sidewaystable}

\section{Discussion and Future Research Directions}

\subsection{From Conceptual to Practical Models for QoE Measurement and Prediction}

In section III, we discussed that QoE modelling involves a complex process of defining relationships between context and QoE parameters with an aim to compute a QoE value on a single scale \cite{brooks2010,mitraicme2011}.  There may be a large number of objective and subjective context parameters that influence users' QoE  in both laboratory and real-life environments. Some of the parameters can be determined while others may be \emph{hidden} i.e., having an indirect affect on users' QoE and thus, are hard to relate and quantify \cite{mitrasac2011}. 

Several researchers \cite{kilkki,marez2007,perkis} proposed conceptual QoE  models that merely classify parameters and their possible relationships.
For example, Perkis, Munkeby and Hillestad \cite{perkis} represented QoE as a tree structure. However, these conceptual models are not practical. These models do not provide unified mechanisms to model, measure and predict QoE. The experts using these conceptual models can merely represent context parameters. They cannot use these models to measure and predict QoE. If they try to extend these conceptual models for performing QoE measurement and prediction, the conceptual representation of parameters may change. For example, parameters represented conceptually as a tree structure may not directly transcend to a mathematical or a statistical model. For example, a regression model. Thus, reducing the scope and benefit of QoE modelling. This necessitates the development of practical models that may benefit experts by providing them a simplified, systematic and a unified approach to model, measure and predict QoE.

We assert that context-aware QoE modelling is a relatively unexplored area and there is a scope to develop novel QoE models. These models should be practical, shareable and reusable across multiple application domains. Unlike \cite{kilkki,marez2007,perkis,laghari12}, these models should also be realistically implementable in systems and applications alike. We believe that probabilistic and ontological models should be extremely beneficial for QoE modelling, measurement and prediction.

\subsection{Methods for QoE Measurement and Prediction}
In section IV, we discussed several methods for QoE measurement and prediction (see table 2 for classification of these methods). We discussed that QoE can be measured using different types of scale and/or by using different units of measurement. For example, a scale of 1 to 5 \cite{mitraicme2011}. QoE can also be measured using by simple {}``yes'' or {}``no'' type questions \cite{menkovski2009}. Typical scales include the ordinal and interval scale \cite{brooks2010,madm}. The transformation of user ratings on the interval scale is difficult, therefore, mostly ordinal scale is used. 

In subjective tests, users select alternatives marked on the ordinal scale  where the distance between alternatives is not fixed. Thus, meaningful results using mathematical operations such as average, standard deviation and ratio cannot be applied \cite{brooks2010,mitraicme2011,Janowski2009,muordinal}. Further, the application of regression-based methods (e.g., \cite{fiedlernetwork}) may also be incorrect. For instance, linear regression requires the residuals to be normally distributed. If users choose only few alternatives on a scale  instead of all, (e.g., only ratings '5' and '4' are selected on the scale of 1 to 5) the error distribution will be asymmetric. Regression techniques only provide the prediction of mean and thus, distribution of the choices is lost \cite{Janowski2009}. As in \cite{Janowski2009,mitraicme2011,muordinal}, we assert that getting a probability distribution instead of computing mean will be correct and may assist the experts in understanding how QoE is distributed based on the underlying test conditions. 

We propose that experts should carefully evaluate the type of data and should carefully choose the statistical techniques they intend to apply. For correct application of any statistical technique, several conditions may have to be verified.  If these conditions are not met, the application of these techniques may be incorrect. Mu \textit{et al.} \cite{muordinal} pointed out that non-parametric statistical techniques should be considered when dealing with subjective tests. We believe that AI-based techniques proposed by Menkovski \emph{et al.} \cite{menkovski2009} and  \cite{mitraicme2011,mitradbn,mitraTMC} can be valuable for QoE measurement and prediction since these techniques can discover relationships between several context and QoE parameters. However, direct application of these techniques can also challenging.

The challenge lies in the fact that these techniques simply classify the alternatives based on some test conditions. For example, consider a case where we have ten user ratings in which six users gave {}``excellent''  and four users gave ``very good''; the AI-based techniques will simply classify alternative as {}``excellent'' with probability 0.60 by ignoring the ratings of other 4 users \cite{mitraicme2011}. To alleviate such problems, Mitra, \AA{}hlund and Zaslavsky \cite{mitraicme2011} used a decision-theoretic approach for QoE measurement and prediction where the authors considered an alternative that ``best describes'' the underlying QoE ratings. This was done via considering all the QoE ratings together to determine a single scalar value. This scalar value was then mapped to the bipolar interval scale to determine the final QoE value.  

In section I and IV, we also highlighted the challenge of QoE measurement on a single scale by considering multiple QoE parameters together \cite{brooks2010,mitraicme2011}. We assert that most of the methods presented in the state-of-the-art concentrate on predicting single QoE parameter independently. Only Gong \textit{et al.} \cite{gongqoemodel} and Mitra, \AA{}hlund and Zaslavsky \cite{mitraicme2011,mitraTMC} presented methods to measure QoE based on multiple QoE parameters. 
 
Finally, we conclude that QoE measurement and prediction over time largely remains an open area of research. We assert that QoE measurement is an evolving process and it should be performed over a period of time (several days, weeks or months depending on the service or application requirements). This may lead to the development of  accurate QoE prediction techniques. In this context, $K^{th}-ordered$ Markov models, HMMs and DBNs  might be valuable as these models can efficiently model users' QoE based on their past ratings as demonstrated by \cite{hossfeldhmm} and \cite{mitradbn}.

\section{Conclusion}
This paper presented a survey of the state-of-the-art research in the area of quality of experience (QoE). We highlighted several challenges associated with QoE modelling, measurement and prediction. We discussed existing methods and highlighted their advantages and shortcomings. This survey also outlined future research directions for QoE modelling, measurement and prediction.

\subsection*{Acknowledgement}
The authors would like to thank Saguna for her valuable feedback and improving the readability of this paper.

\bibliographystyle{plain}
\bibliography{JMD_PHD_REFERENCES,JMD_BHARAT_PHD_REF}

\end{document}